@arxiver{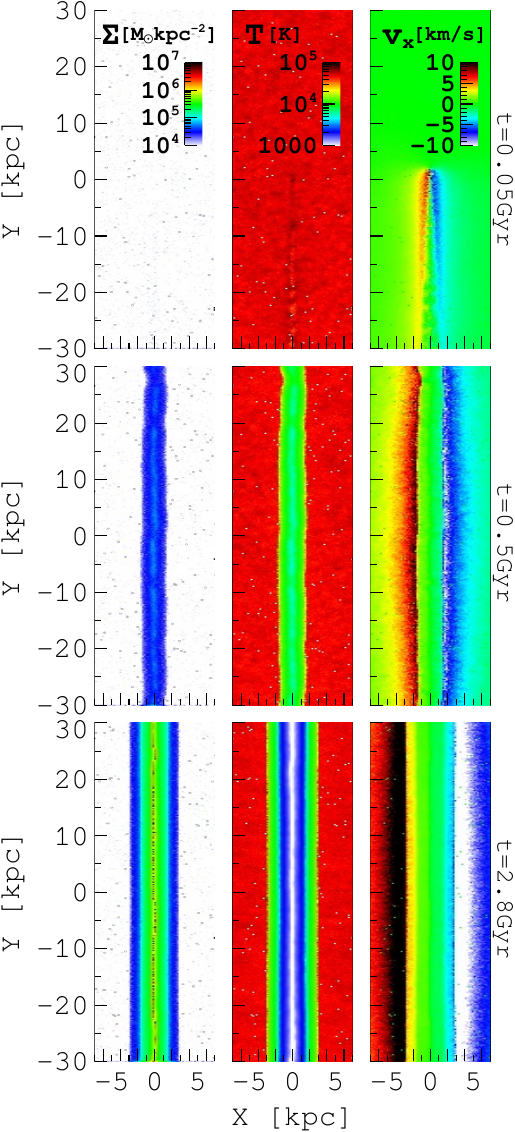,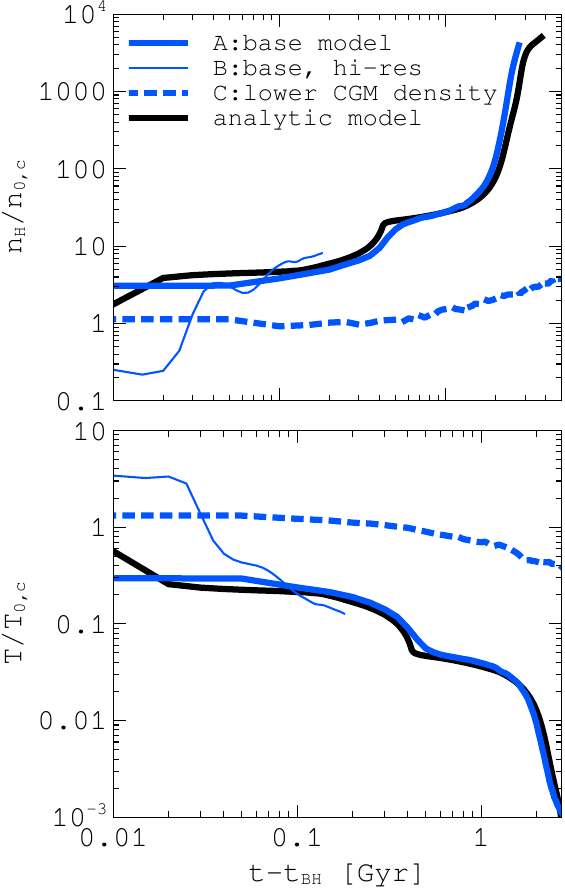,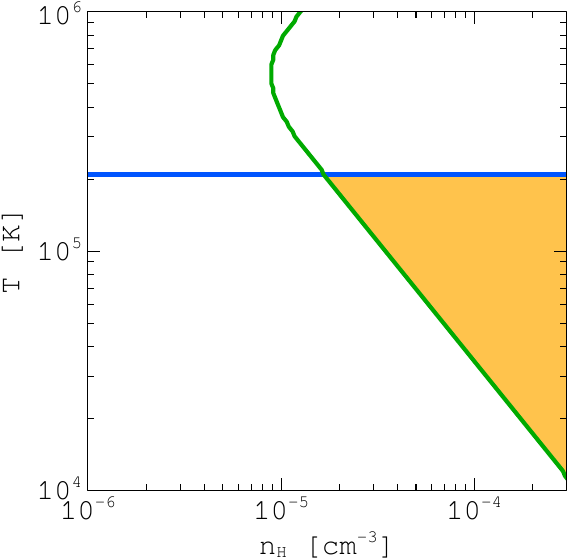}
\documentclass[fleqn,usenatbib]{mnras}

\usepackage{newtxtext,newtxmath}
\usepackage[T1]{fontenc}

\usepackage{graphicx}	% Including figure files
\usepackage[usenames,dvipsnames]{xcolor}
\usepackage{hyperref}
\usepackage[normalem]{ulem}
\usepackage{soul}

\newcommand{\sub}[1]{_{\mathrm{#1}}}
\newcommand{\msun}{M\sub{\sun}}
\defcitealias{vanDokkum2023a}{vD23}

\def\equationautorefname~#1\null{Eq.~(#1)\null}
\def\figureautorefname~#1\null{Fig.~#1\null}
\newcommand{\appref}[1]{\hyperref[#1]{Appendix~\ref{#1}}}

\title[RSMBH triggered filament formation]{Formation of dense filaments induced by runaway supermassive black holes}

\author[G. Ogiya \& D. Nagai]{
Go Ogiya$^{1}$\thanks{E-mail: \url{gogiya@zju.edu.cn} (GO)}, and 
Daisuke Nagai$^{2}$
\\
$^{1}$Institute for Astronomy, School of Physics, Zhejiang University, Hangzhou 310027, China \\
$^{2}$Department of Physics, Yale University, P.O. Box 208121, New Haven, CT 06520, USA \\
}

\date{Accepted XXX. Received YYY; in original form ZZZ}

\pubyear{2023}

\begin{document}
\label{firstpage}
\pagerange{\pageref{firstpage}--\pageref{lastpage}}
\maketitle

\begin{abstract}
A narrow linear object extending $\sim 60$\,kpc from the centre of a galaxy at redshift $z \sim 1$ has recently been discovered and interpreted as shocked gas filament forming stars. The host galaxy presents an irregular morphology, implying recent merger events. Supposing that each of the progenitor galaxies has a central supermassive black hole (SMBH) and the SMBHs are accumulated at the centre of the merger remnant, a fraction of them can be ejected from the galaxy with a high velocity due to interactions between SMBHs. When such a runaway SMBH (RSMBH) passes through the circumgalactic medium (CGM), converging flows are induced along the RSMBH path, and star formation could eventually be ignited. We show that the CGM temperature prior to the RSMBH perturbation should be below the peak temperature in the cooling function to trigger filament formation. While the gas is temporarily heated due to compression, the cooling efficiency increases, and gas accumulation becomes allowed along the path. When the CGM density is sufficiently high, the gas can cool down and develop a dense filament by $z = 1$. The mass and velocity of the RSMBH determine the scale of filament formation. Hydrodynamical simulations validate the analytical expectations. Therefore, we conclude that the perturbation by RSMBHs is a viable channel to form the observed linear object. Using the analytic model validated by simulations, we show that the CGM around the linear object to be warm ($T \la 2 \times 10^5$\,K) and dense ($n \ga 2 \times 10^{-5} (T/2 \times 10^5 \, K)^{-1} \, {\rm cm^{-3}}$).
\end{abstract}

\begin{keywords}
quasars: supermassive black holes -- galaxies: star formation -- ISM: kinematics and dynamics -- methods: numerical
\end{keywords}

%%%%%%%%%%%%%%%%%%%%%%%%%%%%%%%%%%%%%%%%%%%%%%%%%%

%%%%%%%%%%%%%%%%% BODY OF PAPER %%%%%%%%%%%%%%%%%%

\section{Introduction}
\label{sec:intro}

Supermassive black holes (SMBHs) settle at the centre of galaxies, and their mass correlates with the stellar mass of the host galaxies, suggesting that SMBHs co-evolve with galaxies in the process of the hierarchical structure formation in the Universe \citep{Gultekin2009,Kormendy2013}. When galaxies hosting a central SMBH merge, the SMBHs sink to the centre of the merger remnant due to interactions with stars, dark matter, and galactic gas \citep[][and references therein]{Milosavljevic2001,Escala2005,VanWassenhove2014,Ogiya2020}. 

During the close interactions at the galactic centre, a fraction of SMBHs can be ejected from the galaxy \citep{Bekenstein1973} due to the gravitational slingshot effect \citep{Saslaw1974,Volonteri2003,Tanikawa2011} or the gravitational radiation recoil \citep{Baker2006,Kesden2010,Lousto2011} with a velocity of $v > 1,000$\,km/s \citep{Hoffman2007,Campanelli2007}, much larger than the escape velocity of galaxies. Although signatures of such runaway SMBHs (RSMBHs) and their progenitors (i.e., binaries or triplets of SMBHs) are important probes to explore the evolution of SMBHs and thus have been searched for, observational evidence is limited due to the high requirements for observation resolutions \citep[e.g.,][]{Haehnelt2006,Civano2010,Deane2014,Kalfountzou2017}. 

Although the previous efforts of RSMBH searching have been made mainly in the galactic centre, the signature of RSMBHs might be found in the outskirts of galaxies. The strong gravity of RSMBHs produces converging flows along their path in the interstellar medium and the circumgalactic medium (CGM), potentially producing a wake behind the RSMBH \citep{de_la_Fuente_Marcos2008}. Another possible mechanism to form dense filamentary structures is a bow shock produced by the interaction between the gas directly associated with the RSMBH and the ambient gas. Post-shock gas can form a wake of the RSMBH after a cooling catastrophe driven by adiabatic expansion and radiative processes \citep{Saslaw1972}. Recent observations by \citet[][\citetalias{vanDokkum2023a} hereafter]{vanDokkum2023a} discovered a promising candidate for such theoretically predicted structures. They detected a thin, linear object extending 62\,kpc from a galaxy at redshift $z = 0.964$ and interpreted it as a shocked gas filament forming stars. The host galaxy shows a compact irregular morphology and is rich with young stars, suggesting that it experienced a recent merger event \citep[e.g.,][]{Lotz2008,Sazonova2021,Lotz2021}. 

The origin of the observed thin linear object is under debate. \cite{SanchezAlmeida2023a} advocated that the thin linear object is an edge-on bulgeless galaxy based on the position-velocity curve of the object and the Tully-Fisher relation \citep[see also][]{SanchezAlmeida2023b}. As a counterargument, \cite{vanDokkum2023b} presented another observation image that shows that the linear object is connected to the host galaxy and supports the RSMBH-triggered formation scenario. \cite{Chen2023} identified tidal arms/tails that extend $\sim 50$\,kpc and have been formed through consecutive fly-by encounters among three galaxies in the {\texttt ASTRID} cosmological hydrodynamical simulation \citep{Bird2022}. While other observations detected similar collimated gas trails \citep{Yagi2007,Scott2022,Zaritsky2023}, their origin remains unclear. 

The RSMBH-triggered formation scenario for the thin, linear object has rarely been studied on galactic scales ($\ga$\,kpc), while similar processes have been discussed on smaller scales \citep[e.g.,][]{Wallin1996,Li2021,Kitajima2023}. While \cite{de_la_Fuente_Marcos2008} considered the scenario with a simple analytic argument, more detailed studies tailored for the CGM are desired. This paper is the first attempt to test the RSMBH-triggered formation scenario with a combination of analytical argument and high-resolution hydrodynamical simulations. The rest of the paper is structured as follows. In \autoref{sec:model}, we describe the model for studying the response of the CGM to the RSMBH perturbation. The expectations based on analytical considerations are tested using numerical simulations presented in \autoref{sec:simulations}. Finally, we summarise and discuss our findings in \autoref{sec:summary}.

%%%%%%%%%%%%%%%%%%%%%%%%%%%%%%%%%%%%%%%%%%%%%%%%%%
\section{Model Description and Expectation}
\label{sec:model}

%%%%%%%%%%%%%%%%%%%%%%%%%
\subsection{CGM model}
\label{ssec:cgm_model}

We employ a polytrope sphere with an adiabatic index of $\gamma=5/3$ to model the CGM and suppose that radiative cooling is balanced with heating effects to maintain a stable stratified CGM atmosphere (see also \autoref{ssec:cooling_model}). As the polytrope is in the hydrostatic equilibrium state, the gas temperature is determined by the potential depth of the system characterised by the total dynamical mass of the sphere, $M$, and the size of the sphere, $L$. To control the gas density, we introduce another parameter, $f\sub{gas} \equiv M\sub{gas}/M$, where $M\sub{gas}$ is the total gas mass in the sphere and the remaining mass, $(1-f\sub{gas})M$, includes the contributions from dark matter and stars. The parameters, $M, L$ and $f\sub{gas}$, are tuned to have the desired central hydrogen number density ($n\sub{0,c}$) and temperature ($T\sub{0,c}$). The components in the polytrope sphere other than the gas are modelled as a static external potential, and its centre is fixed at the origin of the simulation coordinate system. The numerically computed mass profile is used to calculate the gravity of the external potential. Density and temperature are almost constant at $r \la L/10$, where $r$ is the distance from the centre of the sphere. For simplicity, we consider filament formation in a homogeneous CGM gas environment.

%%%%%%%%%%%%%%%%%%%%%%%%%
\subsection{RSMBH model and the CGM response}
\label{ssec:rsmbh_model}

We assume that the RSMBH path is straight and that fluid elements do not move during interaction with the RSMBH \citep[i.e., impulsive approximation,][]{Binney2008}. These assumptions are tenable when the RSMBH moves in the CGM with a velocity much higher than the sound speed. According to the analytic solution of the hyperbolic encounter between the RSMBH and a fluid element, the kick velocities given to the fluid element in the directions perpendicular and parallel to the RSMBH path are given as
%%%
\begin{equation}
    \Delta v\sub{\perp} = \frac{2v\sub{BH}(b/b\sub{90})}{1+(b/b\sub{90})^2},
        \label{eq:deltav_perp} 
\end{equation}
\begin{equation}
    \Delta v\sub{||} = \frac{2v\sub{BH}}{1+(b/b\sub{90})^2},
        \label{eq:deltav_para}
\end{equation}
%%%
where $v\sub{BH}$ and $b$ indicate the velocity of the RSMBH with respect to the CGM and the minimum distance from the fluid element to the RSMBH path (impact parameter), and $b\sub{90}$ is the impact parameter causing a deflection of 90 degrees and defined as
%%%
\begin{equation}
    b\sub{90} \equiv GM\sub{BH}/v\sub{BH}^2.
        \label{eq:b90} 
\end{equation}
Here, $G$ and $M\sub{BH}$ are the gravitational constant and the RSMBH mass, respectively.

While $\Delta v\sub{\perp}$ produces converging flows that compress the CGM along the RSMBH path \citep{de_la_Fuente_Marcos2008}, $\Delta v\sub{||}$ produces a bow shock that can leave a wake behind the RSMBH \citep{Saslaw1972}. As indicated by \autoref{eq:deltav_perp} and \autoref{eq:deltav_para}, $\Delta v\sub{\perp} \propto b^{-1}$ and $\Delta v\sub{||} \propto b^{-2}$ at $b \gg b\sub{90}$. As such, the former converging flow involves a larger amount of gas than the latter, and hence plays a dominant role in forming dense filaments. Motivated by this consideration, we argue for the formation of dense filaments triggered by converging flows induced by the RSMBH. In the limit of $b \gg b\sub{90}$, \autoref{eq:deltav_perp} is reduced to 
%%%
\begin{equation}
    \Delta v = \frac{2GM\sub{BH}}{b v\sub{BH}}.
        \label{eq:deltav}
\end{equation}
%%%

When the RSMBH gravitationally perturbs the CGM, its pressure is represented as $P = P\sub{thermal} + P\sub{kick}$, where $P\sub{thermal}$ is the thermal pressure of the unperturbed state, and $P\sub{kick}$ is the ram pressure originated by the kick velocity. Thus, we can model it as  
%%%
\begin{equation}
    P = \rho [c\sub{s}^2/\gamma + (\Delta v)^2], 
        \label{eq:pressure}
\end{equation}
%%%
where $\rho$ and $c\sub{s}$ are the density and sound speed of the unperturbed CGM, respectively. Since fluid elements closer to the RSMBH path get kicked more strongly (\autoref{eq:deltav}), the density enhancement can be neglected before the converging flows collide. \autoref{eq:pressure} also implies that the gas temperature at the collision of the converging flows is 
%%%
\begin{equation}
    T\sub{1} = T\sub{0} [1 + \gamma (\Delta v / c\sub{s})^2], 
        \label{eq:temperature}
\end{equation}
%%%
where $T\sub{0}$ ($T\sub{1}$) is the unperturbed (perturbed) CGM temperature. We validate \autoref{eq:temperature} in \autoref{ssec:response} using a hydrodynamical simulation.

%%%%%%%%%%%%%%%%%%%%%%%%%
\subsection{Cooling model and the thermodynamic evolution}
\label{ssec:cooling_model}

%%%%%%%%%%
\begin{figure}
    \begin{center}
        \includegraphics[width=0.45\textwidth]{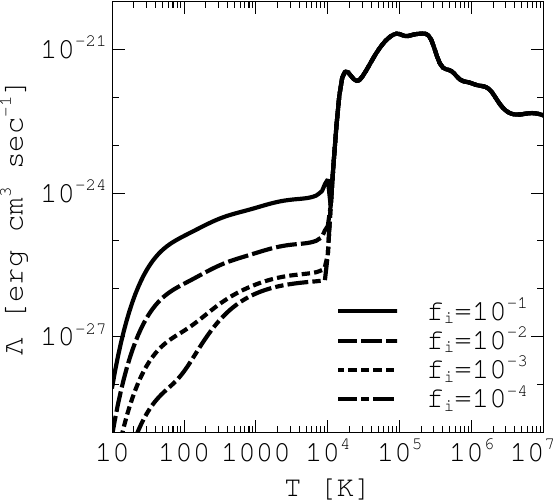}
    \end{center}
    \caption{
        Cooling function. As the slope of the cooling function is negative at $T \geq T\sub{peak} \equiv 2.09 \times 10^5\,$K, the cooling cascade is prevented. At $T \la 10^4$\,K, the cooling efficiency depends on the ionisation fraction, $f\sub{i}$. The assumed $f\sub{i}$ is indicated in the legend.
    \label{fig:cool_func}}
\end{figure}
%%%%%%%%%%

The change in the thermal energy of the fluid is given as 
%%%
\begin{equation}
    \frac{de}{dt} = n\sub{H}\Gamma - n\sub{H}^2 \Lambda(T) - P \frac{dV}{dt},
        \label{eq:ene_eq}
\end{equation}
%%%
where $e$ and $n\sub{H}$ are the internal energy density of the fluid and the hydrogen number density, and $dV/dt$ is the volume change rate. The first and second terms on the right-hand side of \autoref{eq:ene_eq} represent external heating and radiative cooling of the fluid, while the third term is the mechanical work done by the fluid. Here, we assume collisional ionisation equilibrium, where the cooling function ($\Lambda$) depends only on the fluid temperature for a given metallicity, and the heating rate ($\Gamma$) is time-independent. In this study, we employ the cooling function by \cite{Schure2009}, assuming the Solar metallicity (\autoref{fig:cool_func}). At $T \la 10^4$\,K, most atoms become neutral, and the cooling efficiency is dramatically damped. \cite{Schure2009} tabulated the cooling function down to $T = 10$\,K, for the ionisation fraction, $f\sub{i} \equiv n\sub{e}/n\sub{H} = 10^{-1}, 10^{-2}, 10^{-3}$ and $10^{-4}$, where $n\sub{e}$ is the electron number density. In our modelling, $f\sub{i}$ is obtained by solving the Saha ionization equation considering the ground and ionised states of hydrogen at $T \leq 10^{3.8}$\,K. We adopt $f\sub{i} = 10^{-1}$ for $T > 10^{3.8}$\,K and set $f\sub{i} = 10^{-4}$ if lower $f\sub{i}$ is predicted. 

We assume the local cooling-heating balance ($\Gamma = n\sub{0} \Lambda(T\sub{0})$) prior to perturbation by RSMBH, where $n\sub{0}$ is the hydrogen number density of the unperturbed CGM, and the CGM is stable ($dV/dt = 0$). After the RSMBH perturbation, the gas along the RSMBH path needs to cool radiatively to form a dense filament. Suppose that the temperature of the compressed gas increases ($T\sub{1} > T\sub{0}$) while keeping its density and volume constant, as suggested in \autoref{eq:pressure}. The condition required for filament formation is then given by  
%%%
\begin{equation}
    \Lambda(T\sub{1})/\Lambda(T\sub{0}) > 1.
        \label{eq:cool_enhance}
\end{equation}
%%%
Considering the temperature range relevant to the CGM \citep[$10^4 \la T/K \la 10^6$, e.g.,][]{Oppenheimer2016,Tumlinson2017,Lochhaas2020}, we find in \autoref{fig:cool_func} that at $T\sub{0} \geq T\sub{peak} \equiv 2.09 \times 10^5$\,K, \autoref{eq:cool_enhance} will not be satisfied as the cooling function monotonically decreases with temperature. At lower $T\sub{0}$, the condition can be satisfied, making the formation of dense filaments possible. 

\begin{figure}
    \begin{center}
        \includegraphics[width=0.45\textwidth]{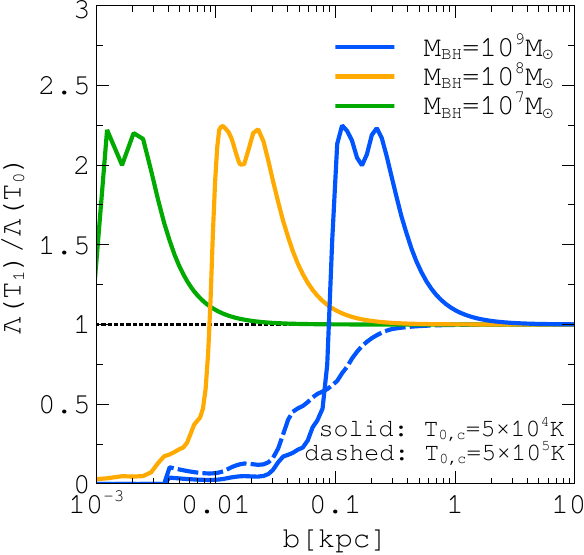}
    \end{center}
    \caption{
        Enhancement of the cooling efficiency as a function of the impact parameter, $b$. In the cases of the solid lines, the gas temperature is $T\sub{0,c} = 5 \times 10^{4}$\,K, while it is $T\sub{0,c} = 5 \times 10^{5}$\,K in the case of the dashed line. The response to the gravitational perturbation by an RSMBH of $M\sub{BH}=10^9$ (blue), $10^8$ (orange), and $10^7 \, \msun$ (green) that runs in the CGM with a velocity of $v\sub{BH} = 1600$\,km/s is considered. The black horizontal dotted line indicates unity for reference. 
        In the cooler CGM model (solid), \autoref{eq:cool_enhance} is satisfied. Thus, the formation of an RSMBH-induced tube surrounding a hot wick is expected, while the scale of the tube formation depends on the RSMBH parameters. The filament formation is unexpected in the hotter CGM (dashed). 
    \label{fig:cool_enhance}}
\end{figure}

As discussed in \autoref{ssec:rsmbh_model}, the amplitude of the kick velocity and temperature enhancement by the RSMBH perturbation depend on $M\sub{BH}, v\sub{BH}$, and $b$. In \autoref{fig:cool_enhance}, we consider the condition of \autoref{eq:cool_enhance}, varying the RSMBH parameters as well as the CGM temperature of the unperturbed state, $T\sub{0}$. We adopt the fiducial central temperature of $T\sub{0,c} = 5 \times 10^{4}$\,K $< T\sub{peak}$ (solid lines) and check how our results change for a higher temperature of $T\sub{0,c} = 5 \times 10^{5}$\,K $> T\sub{peak}$ (dashed lines). \autoref{eq:cool_enhance} indicates that filament formation is possible in the former, while it is unexpected in the latter, allowing us to narrow the CGM parameter space to explore. In the cases of the cooler CGM model, the RSMBH mass, $M\sub{BH}$, is varied, while it is $M\sub{BH}=10^9\,\msun$ in the hotter CGM case. In all cases, the RSMBH velocity is fixed ($v\sub{BH} = 1600$\,km/s). Note that \autoref{eq:deltav} indicates that the impact of increasing $M\sub{BH}$ by a factor of $A$ is the same as that of decreasing $v\sub{BH}$ by the same factor. 

%%%%%%%%%%
%%%%%%%%%%

We present the results of the analysis in \autoref{fig:cool_enhance}. At small $b$, a high kick velocity is induced (\autoref{eq:deltav}), and the gas temperature is significantly increased ($T \geq T\sub{peak}$). As a consequence, the cooling efficiency decreases in all cases. When the initial CGM temperature is lower than the peak temperature, there exists a $b$-range of the enhanced cooling efficiency (solid), and the range depends on $M\sub{BH}$, i.e., the larger $M\sub{BH}$, the larger $b$-range of the enhanced cooling efficiency. Thus, a cool gas tube surrounding a hot, thinner wick can be formed in the cooler CGM models. Considering the hotter CGM model (dashed), the cooling efficiency is reduced at all $b$-ranges, preventing filament formation. When \autoref{eq:cool_enhance} is satisfied, the cooling cascade can occur along the RSMBH path. 

Another condition required for forming dense filaments is that the gas should be cooled down enough by the age of the observed Universe:
\begin{equation}
    \tau\sub{cool} < t\sub{obs},
        \label{eq:tcool_condition}
\end{equation}
where $t\sub{obs} \sim 5.9$\,Gyr for the observed filament. We then estimate the net cooling timescale, $\tau\sub{cool}$, by integrating \autoref{eq:ene_eq} from $T = T\sub{0}$ to the minimum temperature of $T = 10$\,K in the cooling function table.

%%%%%%%%%%%%%%%%%%%%%%%%%%%%%%%%%%%%%%%%%%%%%%%%%%
\section{Simulations}
\label{sec:simulations}

%%%%%%%%%%%%%%%%%%%%%%%%%
\subsection{Simulation setup}
\label{ssec:sim_setup}

The hydrodynamical evolution of the polytrope is simulated using a publicly available code, \texttt{GIZMO} \citep{Hopkins2015}\footnote{\url{https://bitbucket.org/phopkins/gizmo-public/src/master/}}, in which various particle-based computational fluid dynamics schemes are available. We employ the default setting of the hydrodynamics solvers in \texttt{GIZMO}. 32 neighbour particles weighted with the cubic spline kernel are used to estimate the density and solve the hydrodynamic equations. Magnetic fields are not considered. In each simulation, we employ 20,000,000 particles. 

The three-dimensional position vector of each particle is drawn using the acceptance-rejection sampling method \citep{Press2002} based on the density profile of the polytrope, while they initially have zero velocity. The thermal energy is assigned to each particle based on the drawn radius and the temperature profile of the polytrope. While the simulated system is spherical, we concentrate the numerical resource in the region of our particular interest, a cylinder around the path of the RSMBH ($y$-axis in most cases, see below). Specifically, half of the particles (10,000,000) are located within a cylinder of a diameter, $d = 2L/15$, and a length, $l = 2L/3$. Thus, the mass resolution of particles in the cylinder is increased by $\approx 84$, compared to that of particles outside the cylinder. This creates a resolution gap at the interface of the high-resolution cylinder and can cause artificial surface tension. After some experiments, we find that the pressure-entropy formalism of smoothed particle hydrodynamics \citep[PSPH,][]{Saitoh2013,Hopkins2013} minimises such artefacts. Therefore, PSPH is employed in this study. Before introducing an RSMBH, the polytrope sphere is relaxed in isolation, and the initial configuration (e.g., density and temperature profiles and cylinder shape) of the polytrope is well-kept. At the beginning of the isolated evolution, ripples in the spatial and energy distributions of particles are excited as a result of the Poisson noise and the resolution gap. The fluctuation in the gas density and temperature is less than five percent of those in the analytical polytrope model, and we employ a snapshot of the well-relaxed state as the initial condition of the simulations with an RSMBH.

%%%%%%%%%%
\begin{figure}
    \begin{center}
        \includegraphics[width=0.45\textwidth]{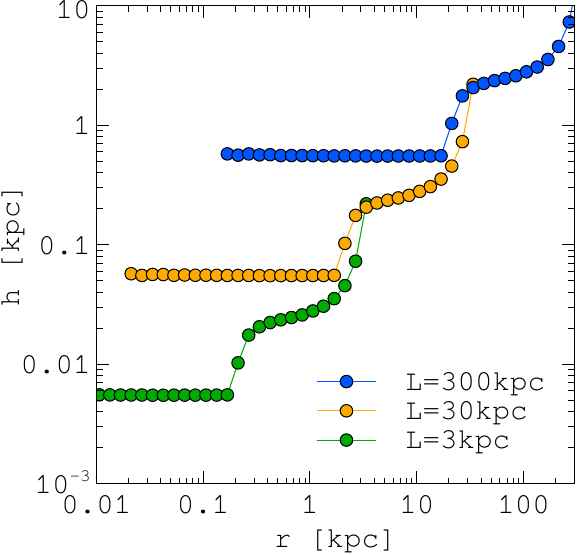}
    \end{center}
    \caption{
        Smoothing length of particles, $h$, as a function of the distance from the centre of the polytrope, $r$. The first snapshot of the simulations is analysed. The blue, orange, and green lines represent the model with the radius of the sphere, $L = 300$, 30 and 3\,kpc, respectively. As a high-resolution cylinder is placed in the centre, $h$ drops by $\sim 4$ at $r = L/10$. 
    \label{fig:smoothing_length}}
\end{figure}
%%%%%%%%%%

The smoothing length of gas particles ($h$) corresponds to the spatial resolution of smoothed particle hydrodynamics simulations. In \autoref{fig:smoothing_length}, we show $h$ as a function of the distance from the centre of the polytrope, $r$. As $h$ of a gas particle is determined as the radius of a sphere containing 32 neighbour particles measured from the particle, it gets smaller toward the centre of the polytrope, where the gas density (i.e., particle number density) gets higher. The sudden drop in $h$ at $r \la L/10$ represents the higher resolution cylinder that increases spatial resolution by $\sim 4$. 

It is important to accurately simulate the scales of enhanced cooling efficiency, as shown in \autoref{fig:cool_enhance}. For example, when the black hole mass is $M\sub{BH}=10^8\,\msun$, the smoothing length $h$ should be less than $\sim 0.2$\,kpc. It is worth noting that we refer to the initial smoothing length, as the cooling efficiency is enhanced immediately after the RSMBH perturbation, although $h$ can become smaller as the particles accumulate during the evolution. In simulations with insufficient resolution, the formation of dense filaments is artificially suppressed, as illustrated in \autoref{fig:tube_evo_mbh}.
\footnote{As discussed in \autoref{ssec:rsmbh_model}, the formation of the filament triggered by the bow shock could be efficient on the scale of the impact parameter of the strong encounter $b\sub{90}$. Having $v\sub{BH} = 1600 \, {\rm km/s}$, we get $b\sub{90} \sim 1.6 (M\sub{BH}/10^9 \msun) \, {\rm pc}$, implying that higher spatial resolutions are required to investigate the mechanism.}

The RSMBH is modelled with an external potential of a point mass smoothed with a Plummer softening parameter of $\epsilon = L/10^4$. The potential centre is initially set at $(x,y,z) = (x\sub{BH}, -4L/15, 0)$ and moves to the $y$-direction with a velocity, $v\sub{BH} = v\sub{y} = 1600$\,km/s (\citetalias{vanDokkum2023a}), while velocities in the $x$- and $z$-directions are zero in the simulations. Here, $x\sub{BH}$ is the offset from the $y$-axis to the $x$-direction. The assumption of a straight orbit and the use of analytic potential are justified under the condition that $v\sub{BH}$ is much greater than the escape velocity of the host galaxy of the CGM, which is the case for our problem. 

As discussed in \autoref{ssec:cooling_model}, the perturbation by the RSMBH may break the cooling-heating balance assumed in the unperturbed CGM. In our simulations, we turn on radiative cooling for particles that satisfy the following two conditions:  
%%%
\begin{equation}
    n\sub{H} \Lambda(T) \geq \alpha n\sub{0,c} \Lambda(T\sub{0,c}) \wedge n\sub{H} \geq \beta n\sub{0,c}, 
        \label{eq:cooling_switch}
\end{equation}
%%%
where $\alpha$ and $\beta$ are free parameters. We employ $\alpha = \beta = 1.2$ in all simulations, and the results are insensitive to the choice of parameters. Once radiative cooling is turned on, it works with $\Gamma = 0$. While this model of radiative cooling can be improved, we employed it in the current study for simplicity, and studying filament formation under the self-consistent modelling of radiative cooling is left for future projects. To properly follow the cooling evolution of the gas in simulations, the instantaneous cooling timescale, $t\sub{cool} \equiv (3/2) k\sub{B} T/[\mu f\sub{H} n\sub{H} \Lambda(T)]$, must be well resolved, where $k\sub{B}$ is the Boltzmann constant, $\mu$ is the mean molecular weight, $f\sub{H}$ is the hydrogen mass fraction, and $T$ is the temperature at a given time, respectively. In the simulations, we employ an explicit first-order scheme to solve \autoref{eq:ene_eq} with the cooling timestep, $\Delta t\sub{cool} \leq t\sub{cool}/10$ \citep{Townsend2009}. 

%%%%%
\begin{table}
\begin{center}
\caption{Summary of the simulation parameters. In all simulations, the initial temperature at the polytrope centre is $T\sub{0,c} = 5 \times 10^4$\,K, and the velocity of the RSMBH with respect to the polytrope is 1,600\,km/s. The RSMBH path corresponds to the $y$-axis, i.e., $x\sub{BH}=0$. 
Column 
(1) Simulation ID. 
(2) Radius of the polytrope. 
(3) Central hydrogen number density.  
(4) Mass of the RSMBH. 
\label{tab:params}}
\begin{tabular}{cccc}
\hline 
(1)         & (2)        & (3)                   & (4)                 \\
ID          & $L$ [kpc]  & $n\sub{0,c}$ [${\rm cm}^{-3}$] & $M\sub{BH} [\msun]$ \\
\hline
A           & 300        & $10^{-4}$             & $10^9$              \\
B           & 30         & $10^{-4}$             & $10^9$              \\
C           & 300        & $10^{-6}$             & $10^9$              \\
D           & 30         & $10^{-4}$             & $10^8$              \\
E           & 300        & $10^{-4}$             & $10^8$              \\
F           & 3          & $10^{-4}$             & $10^7$              \\
G           & 30         & $10^{-4}$             & $10^7$              \\
\hline
\end{tabular}
\end{center}
\end{table}
%%%%%

In our model, the CGM is modelled with a spherically-symmetric polytrope. To investigate how the symmetry of the system can affect the results, we performed simulations varying $x\sub{BH}$. We found that the system's symmetry hardly affects the simulation results. Thus, we employ $x\sub{BH} = 0$, and the RSMBH path in our simulations corresponds to the $y$-axis. \autoref{tab:params} summarises simulation parameters. The simulation results are numerically converged.

%%%%%%%%%%%%%%%%%%%%%%%%%
\subsection{CGM response in the early phase}
\label{ssec:response}

%%%%%%%%%%
\begin{figure}
    \begin{center}
        \includegraphics[width=0.45\textwidth]{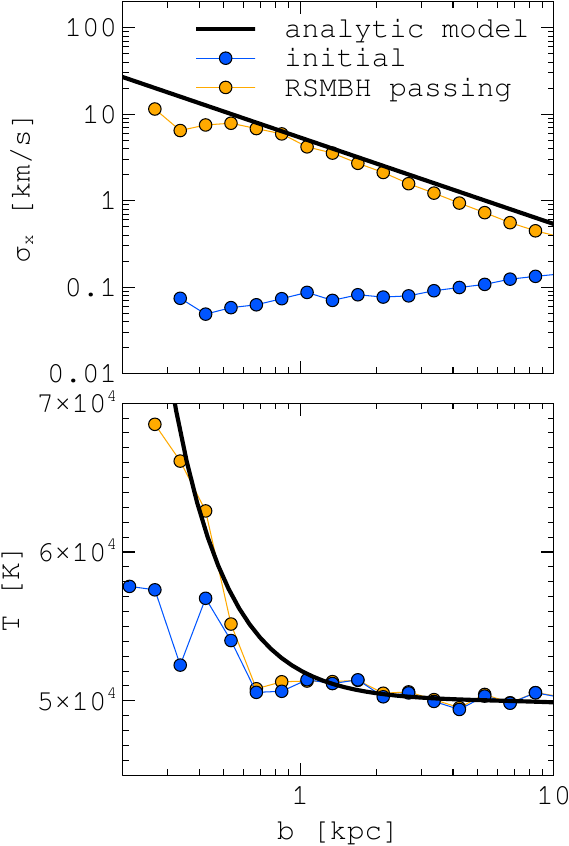}
    \end{center}
    \caption{
        Response of the polytrope to the gravitational perturbation by the RSMBH. Particles at $|y| \leq 0.5$\,kpc in run-A are considered. The impact parameter, $b$, indicates the distance from the $y$-axis. 
        ({\it Upper}) Velocity inducement in the CGM. The profile of the $x$-component of velocity dispersion, $\sigma\sub{x}$, is shown, as the mean velocity is zero because of the symmetry. 
        ({\it Lower}) The temperature profile of the CGM. 
        The black lines in the upper and lower panels are the analytical predictions (\autoref{eq:deltav} and \autoref{eq:temperature}). Blue and orange lines show the initial condition of the simulation and the snapshot at which the RSMBH passes through the centre of the polytrope sphere. The analytical model explains the simulation result. 
    \label{fig:response}}
\end{figure}
%%%%%%%%%%

Here, we test the analytical expectation of the induced velocity (\autoref{eq:deltav}) and the temperature response (\autoref{eq:temperature}) of the CGM to the RSMBH passage. \autoref{fig:response} compares the analytical expectation (black) to the results from run-A. While the blue line shows the initial condition of the simulation, the orange line is the snapshot at which the RSMBH is passing through the centre of the polytrope sphere. We find in the upper panel that the amplitude of the induced velocity is consistent with the analytical expectation, and the RSMBH creates the converging flows along its path. As the converging flows compress the gas, the gas temperature around the RSMBH path is raised, and the temperature enhancement is consistent with the analytical prediction, as shown in the lower panel. Therefore, the cooling efficiency is enhanced by gas compression, which is the first condition for filament formation (\autoref{eq:cool_enhance}), and it is expected that a filament will form when the second condition (\autoref{eq:tcool_condition}) is satisfied.

%%%%%%%%%%%%%%%%%%%%%%%%%
\subsection{Filament formation}
\label{ssec:filament_formation}

%%%%%%%%%%
\begin{figure}
    \begin{center}
        \includegraphics[width=0.42\textwidth]{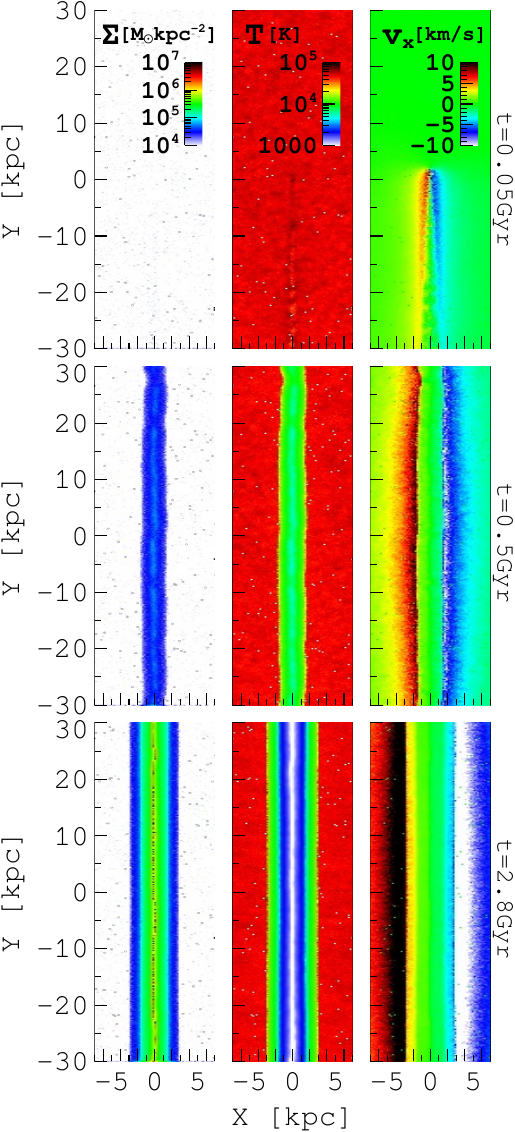}
    \end{center}
    \caption{
        Formation of the condensed gas filament in run-A.  
        ({\it Left}) Column density in $\msun/{\rm kpc}^{2}$. 
        ({\it Centre}) Temperature in K. 
        ({\it Right}) Velocity in the $x$-direction in km/s. 
        From top to bottom, the snapshots at $t = 0.05, 0.5$ and 2.8\,Gyr are demonstrated. Particles at $|z| \leq 1$\,kpc are considered. 
        The RSMBH induces converging flows, and the gas is compressed (top), enhancing the cooling efficiency. This allows further accumulation of the gas and formation of the dense gas filament with a length of $> 60$\,kpc (bottom). 
    \label{fig:run-a}}
\end{figure}
%%%%%%%%%%

In \autoref{fig:run-a}, we demonstrate the distribution of column density (left), temperature (centre), and velocity in the $x$-direction (right) in run-A. The simulation model corresponds to the solid blue line in \autoref{fig:cool_enhance}. The particles at $|z| \leq 1$\, kpc are considered in this analysis. From the top to the bottom panels, the time evolution is depicted. When the RSMBH is passing through the polytrope centre (top row), the impact of the RSMBH perturbation on the $v\sub{x}$-map is evident, while the temperature along the RSMBH path ($y$-axis) is slightly raised. On the other hand, the column density remains almost constant, as set by the polytrope model. Converging flows are induced along the RSMBH path with the amplitude of the kick velocity, consistent with the analytical prediction (\autoref{fig:response}, upper). As the cooling efficiency is enhanced by gas compression induced by the converging flows, the gas starts cooling along the path. Density (temperature) increases (decreases) by a factor of $\sim 10$ in a short time ($\sim 10^7$\,yr). Radiative cooling along the path makes the converging flow stronger. Then, gas condensation is delayed as the efficiency of radiative cooling dramatically drops at $T \la 10^4$\,K (middle row). Even with a reduced cooling efficiency, gas accumulation is not halted. At $t = 2.8$\,Gyr (bottom row), the column density is enhanced by $\sim$ 100 from the initial density. Tracking the subsequent evolution of the CGM is numerically difficult as dense parts of the RSMBH path collapse due to a cooling catastrophe, indicating the formation of a dense filament that can potentially form stars. 

%%%%%%%%%%
\begin{figure}
    \begin{center}
        \includegraphics[width=0.42\textwidth]{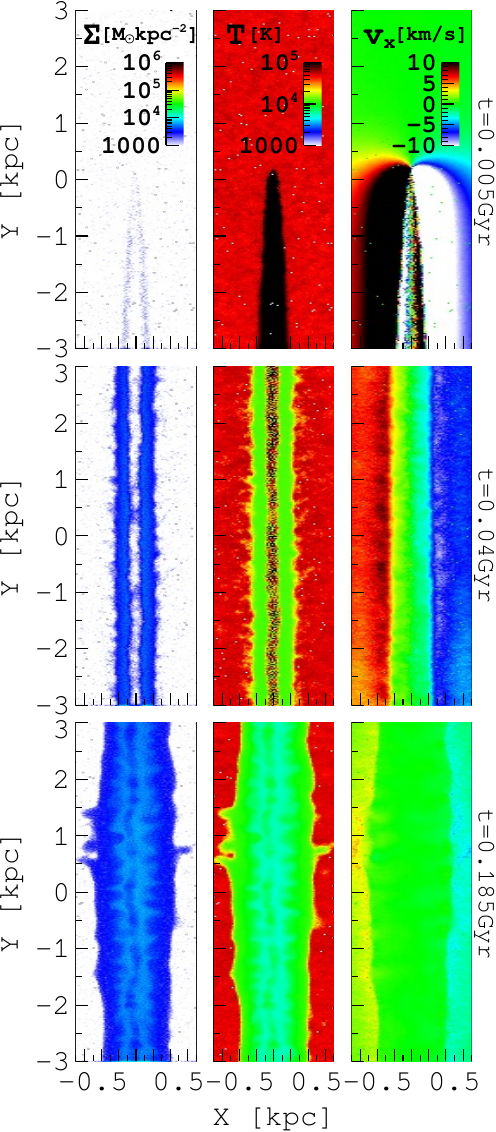}
    \end{center}
    \caption{
        Same as \autoref{fig:run-a}, but the size of the polytrope is smaller by a factor of 10, making the spatial resolution in the simulation ten times higher (run-B). From top to bottom, the snapshots at $t = 0.005, 0.04$ and 0.185\,Gyr are demonstrated. Particles at $|z| \leq 0.1$\,kpc are considered. 
        As predicted by the analytical model in \autoref{ssec:cooling_model}, a hot and diffuse wick has been created and is surrounded by a cold and dense tube (middle). The wick is mixed with the tube in a short timescale (bottom). 
    \label{fig:run-b}}
\end{figure}
%%%%%%%%%%

An intriguing prediction by the analytical model in \autoref{sec:model} is that the RSMBH perturbation also triggers the formation of a hot and diffuse wick inside the cold and dense tube (\autoref{fig:cool_enhance}). If the high-pressure wick persists for a long time, the accumulation of the gas onto the RSMBH path can be prevented. When considering the model of run-A, the wick is expected to be formed at $b \la 0.1$\,kpc. However, run-A lacks spatial resolutions to resolve such small scales. In run-B, we increase the spatial resolution by employing a smaller polytrope, while the CGM and RSMBH parameters are identical to those in run-A to investigate how long the wick lasts. The results from run-B are demonstrated in \autoref{fig:run-b}. The amplitudes of the induced velocity and the temperature enhancement when the RSMBH passes the CGM are larger than those in run-A since the small scale is resolved (top row). As the smaller scales are more significantly heated, a hot and diffuse wick surrounded by a cold and dense tube has been formed, as expected (middle row). The wick has been mixed with the surrounding colder tube in 0.2\,Gyr (bottom row). The results in \autoref{fig:run-b} hardly change even if the radiative cooling is turned off at high temperature, $T > T\sub{peak}$. Therefore, the wick hardly affects the long-time development of the dense gas filament shown in \autoref{fig:run-a}. 

%%%%%%%%%%
\begin{figure}
    \begin{center}
        \includegraphics[width=0.42\textwidth]{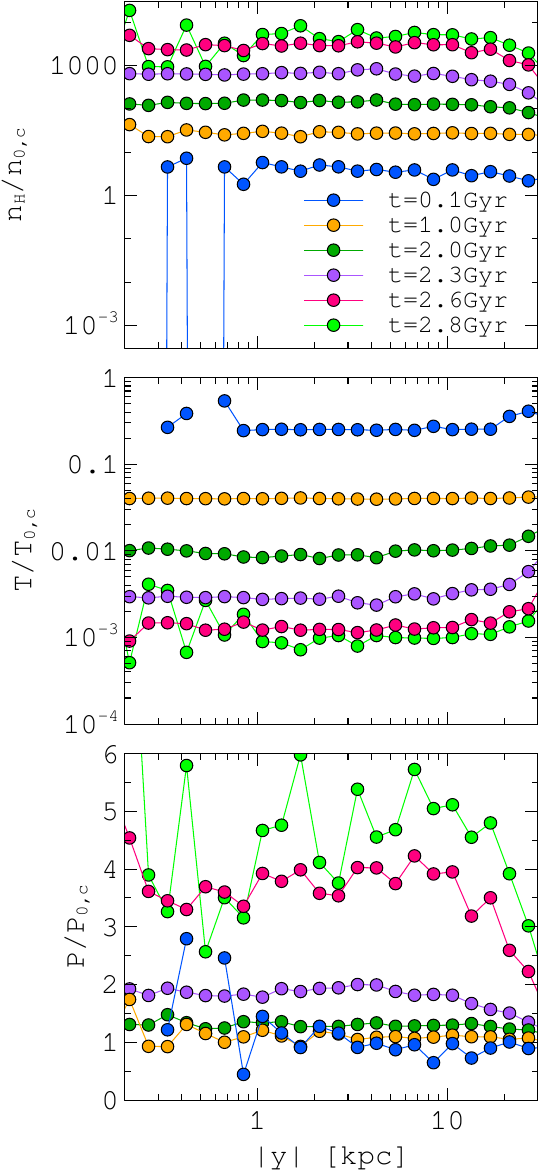}
    \end{center}
    \caption{
        Evolution of the hydrogen number density, $n\sub{H}$ (top), gas temperature, $T$ (middle), and pressure, $P$ (bottom) in run-A. The quantities are scaled with the analytical values measured at the centre of the polytrope (indicated by the subscript '0,c'). Particles within a cylinder along the $y$-axis with a radius of $R = 0.1$\,kpc are considered. Blue, orange, green, purple, red, and light-green lines show the snapshots at $t = 0.1, 1.0, 2.0, 2.3, 2.6$, and 2.8\,Gyr. The density (temperature) increases (decreases) with time and does not strongly depend on $y$, indicating the development of a dense and cold gas filament. The filament forms through the isobaric evolution, approximately.
    \label{fig:profiles}}
\end{figure}
%%%%%%%%%%

\autoref{fig:profiles} presents the profiles of the gas density (top), temperature (middle), and pressure (bottom) in run-A. Particles within a cylinder with a radius of $R = 0.1$\,kpc along the RSMBH path ($y$-axis) are included in the analysis. We compute the density at given $y$ by dividing the total mass of particles within the volume element by its volume. Similarly, we compute the temperature from the average internal energy at given $y$. Since $R$ is smaller than $h$ (the smoothing length) at $t=0$, the simulation does not properly resolve the cylinder of interest. Despite this fact, this scheme reproduces the central density ($n\sub{0,c}$) and temperature ($T\sub{0,c}$) of the analytical polytrope model at $t=0$. At later phases, $h$ decreases with time as the density increases and the cylinder is well resolved. Blue, orange, green, purple, red, and light-green lines represent the snapshots at $t = 0.1, 1.0, 2.0, 2.3, 2.6$, and 2.8\,Gyr, respectively. The density and temperature are almost independent of $y$. The density increases as the gas cools down with time, forming a dense, cold gas filament. By $t = 2.8$\,Gyr, the density and temperature are enhanced and reduced by a factor of $\sim 10^3$, respectively, while keeping the pressure constant and increasing by a factor of only $\sim$~5 at the last phase of the simulation (purple, red and light-green). As the gas accumulates, its self-gravity becomes dominant and plays a role in enhancing the thermal pressure of the developing filament. As the change in pressure is much smaller than those in density and temperature, the isobaric process is a reasonable approximation to model the evolution of the filament.

%%%%%%%%%%%%%%%%%%%%%%%%%
\subsection{Dependence of filament formation on CGM and RSMBH parameters}
\label{ssec:dependence_on_cgm_and_rsmbh}

In this subsection, we study how filament formation depends on the parameters of the CGM and RSMBH models. The simulation model contains four physical parameters to vary: the density and temperature of the CGM, $n\sub{0,c}$ and $T\sub{0,c}$ respectively, and the mass and velocity of the RSMBH, $v\sub{BH}$ and $M\sub{BH}$. Here, only $n\sub{0,c}$ and $M\sub{BH}$ are varied, while $T\sub{0,c}$ and $v\sub{BH}$ are fixed to those employed in runs-A and -B ($T\sub{0,c} = 5 \times 10^4$\,K $< T\sub{peak}$ and $v\sub{BH} = 1,600$\,km/s). As discussed in \autoref{fig:cool_enhance}, the RSMBH-induced filament formation is not expected when the CGM is hotter than $T\sub{peak}$ prior to RSMBH perturbation, as cooling efficiency is reduced by gas compression caused by the RSMBH passage. When the CGM is cooler than $T\sub{peak}$, after the RSMBH perturbation, \autoref{eq:cool_enhance} is satisfied at certain scales, and the thermal evolution of the gas would be qualitatively the same, i.e., radiative catastrophe followed by temporal gas heating due to gas compression. Thus, we fix $T\sub{0,c}$. The dependence of the CGM response on $v\sub{BH}$ degenerates with that on $M\sub{BH}$, as suggested by \autoref{eq:deltav}. Our simulations confirm this degeneracy. Thus, we vary $M\sub{BH}$ while $v\sub{BH}$ is fixed.

Using the analytic model described in \autoref{ssec:cooling_model}, we derive the evolution of the gas density and temperature to compare to the simulation results. As shown in \autoref{fig:profiles}, the cylinder evolves isobarically. \autoref{eq:ene_eq} suggests that to maintain the isobaric process, 
%%%
\begin{equation}
    -n\sub{H}^2 \Lambda(T) dt = (1/3)P dV,
        \label{eq:isobaric_condition}
\end{equation}
%%%
should be satisfied under the assumption of $\Gamma = 0$. Here, the factor of 1/3 appears, as the gas compression happens only in the radial direction of the cylinder. We numerically integrate \autoref{eq:isobaric_condition} with the time-stepping of $dt = t\sub{cool}/10$ and update $n\sub{H}$ and $T$ with the constraint of $P = P(t=0)$. 

%%%%%%%%%%
\begin{figure}
    \begin{center}
        \includegraphics[width=0.45\textwidth]{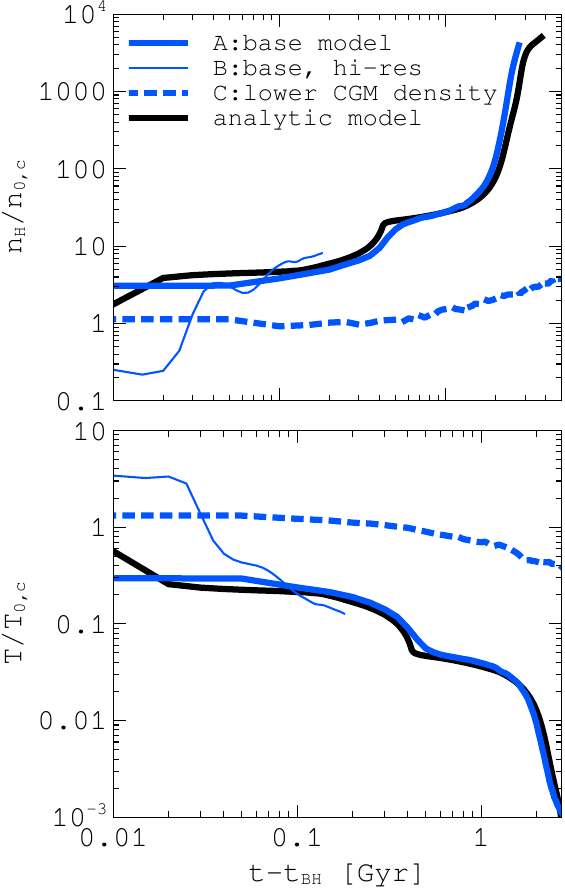}
    \end{center}
    \caption{
        Evolution of the density (upper) and temperature (lower) of the gas in a cylinder with a radius of $R=0.1$\,kpc since the time, $t\sub{BH}$, at which the RSMBH passes through the polytrope centre. Runs-A and -B follow (thick and thin solid blue) the expected evolutional track (black). The evolution is delayed in run-C (blue dotted) as the cooling efficiency is lower because of the lower CGM density. 
    \label{fig:tube_evo_ncgm}}
\end{figure}
%%%%%%%%%%

In \autoref{fig:tube_evo_ncgm}, we present the evolution of the gas density (upper) and temperature (lower) within a cylinder along the $y$-axis. The radius of the cylinder is 0.1\,kpc in the analysis. We find that the results from run-A (thick solid blue) agree with the analytic predictions (black). The density and temperature of the gas have been enhanced and reduced by more than $10^3$ in the evolution of a few Gyr, indicating the development of a dense and thin filament. The formation of the hot and diffuse wick inside the cool and dense tube is found in run-B, and the wick is mixed with the tube in a short timescale, as demonstrated in \autoref{fig:run-b}. Consequently, the evolutional track in run-B approaches the analytic prediction. The CGM density is lowered by a factor of 100 in run-C. As a result, the cooling timescale gets longer, and the development of the filament is delayed (dotted blue). The results imply that the condition of \autoref{eq:tcool_condition} sets the lower limit of the CGM density to form the observed filamentary structure. 

Then, we consider how the CGM response depends on $M\sub{BH}$. According to \autoref{fig:cool_enhance}, higher spatial resolutions are required to resolve the formation of filaments triggered by RSMBHs of smaller $M\sub{BH}$. For instance, the spatial resolution of $\la 0.1 (0.01)$\,kpc is required to resolve the formation of filaments induced by the RSMBH of $M\sub{BH} = 10^8 (10^7) \, \msun$. To achieve such a high resolution (see \autoref{fig:smoothing_length}), we employ the polytrope of $L = 30 (3)$\,kpc in run-D (run-F), studying the response of the CGM to the gravitational perturbation of the RSMBH with $M\sub{BH} = 10^8 (10^7) \, \msun$. Although higher spatial resolution can also be obtained by increasing the number of particles, $N$, it is unfeasible to achieve the required resolution with a larger polytrope, as the dependence of $h$ on $N$ is relatively weak ($h \propto N^{-1/3}$). For comparison, lower resolutions are employed in runs-E and -G, while the CGM and RSMBH parameters are the same as those in runs-D and -F. 

\autoref{eq:deltav} implies that a fluid element located at $b$ from the RSMBH path will reach the path at $t = b / \Delta v = b^2 v\sub{BH}/(2GM\sub{BH})$. Thus, one may estimate the width of the compressed gas at $t$ with the impact parameter, 
%%%
\begin{equation}
    b\sub{comp} = (2GM\sub{BH}t/v\sub{BH})^{1/2}. 
        \label{eq:bcomp}
\end{equation}
%%%
The response of the gas, such as the enhancement of pressure and temperature to the RSMBH with different masses, is expected to be the same at $B b\sub{comp}$ for given $v\sub{BH}$ and $t$, where $B$ is a given constant factor. Therefore, we consider particles within a cylinder with a radius of $R = 0.1(M\sub{BH}/10^9\,\msun)^{1/2}$\,kpc along the $y$-axis in the analysis. 

%%%%%%%%%%
\begin{figure}
    \begin{center}
        \includegraphics[width=0.45\textwidth]{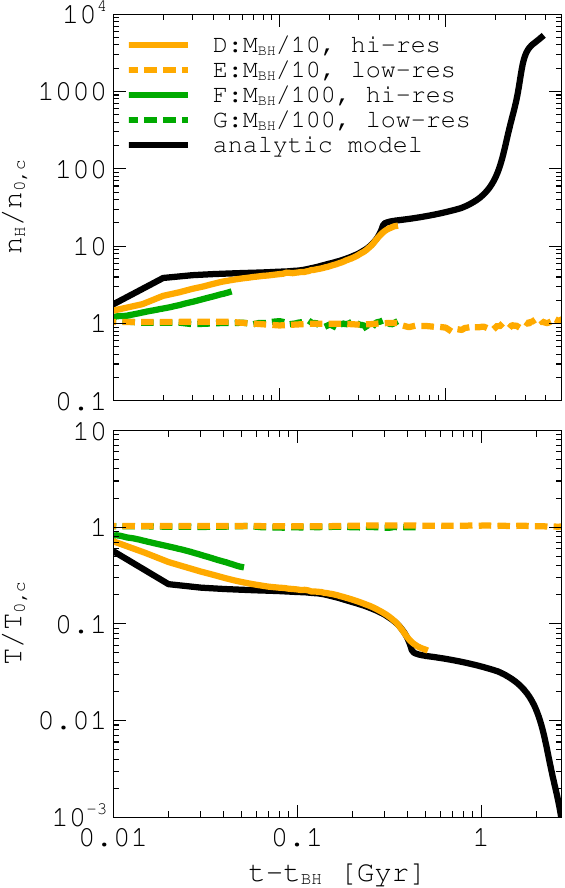}
    \end{center}
    \caption{
        Evolution of the density (upper) and temperature (lower) of the gas in a cylinder with a radius of $R=0.1(M\sub{BH}/10^9\,\msun)^{1/2}$\,kpc since the time, $t\sub{BH}$, at which the RSMBH passes through the polytrope centre. To reduce the Poisson noise, $R=0.1$\,kpc is employed in analysing run-E. Simulations with sufficient resolutions (runs-D and -F, solid orange and green) reasonably agree with the analytic prediction (black), independent of $M\sub{BH}$. The formation of the dense filament is artificially suppressed when numerical resolutions are insufficient (runs-E and -G, dotted). 
    \label{fig:tube_evo_mbh}}
\end{figure}
%%%%%%%%%%

We study the dependence of filament formation on $M\sub{BH}$ in \autoref{fig:tube_evo_mbh}. Here, we vary not only $M\sub{BH}$ but also the radius of the polytrope, $L$, to control spatial resolution in the simulation. $L$ also sets the timescale so that the simulation can follow the evolution of the gas. When the passage of the RSMBH perturbs the gas and induces radiative cooling, the core of the polytrope ($r = L/10$) starts to collapse at a few sound crossing times. For the polytropic CGM model with the same $T\sub{0,c}$ and different $L$ values, the sound crossing time is proportional to $L$.

In simulations with sufficient resolutions (runs-D and -F, solid orange and green), the evolution of the density and temperature of the gas of the cylinder reasonably agrees with the analytic predictions (black) independent of $M\sub{BH}$, while the development of the dense filament is still on the way. As discussed above, the timescale allowed to follow is shorter in these runs than in run-A because of the smaller $L$, and following the long-term response of the CGM to RSMBHs with a smaller $M\sub{BH}$ is left for future studies. Nevertheless, the 'seeds' of a dense filament will trace the track derived by the analytic model in the long-term evolution, as with run-A, for the following two reasons. First, the CGM density and temperature in runs-D and -F are the same as those in run-A. Second, the self-similar response of the CGM to the RSMBH perturbation is expected, as shown in \autoref{fig:cool_enhance}.

Runs-E and -G (orange and green dotted) have the same CGM and RSMBH parameters as those in runs-D and -F, respectively. However, the density and temperature remain constant throughout the simulation. These simulations employ a larger $L$, which allows one to follow the gas evolution for longer while the spatial resolution gets lower. Due to insufficient resolutions, filament formation is artificially suppressed in these simulations.

%%%%%%%%%%%%%%%%%%%%%%%%%
\subsection{CGM conditions for the RSMBH-induced filament formation}
\label{ssec:condition_filament_formation}

%%%%%%%%%%
\begin{figure}
    \begin{center}
        \includegraphics[width=0.45\textwidth]{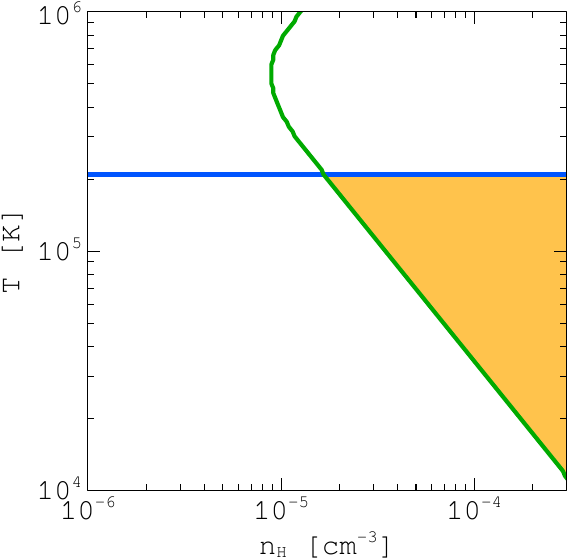}
    \end{center}
    \caption{
        CGM parameter space to form dense gas filaments. The blue line corresponds to the peak temperature in the cooling function, $T\sub{peak}$, below which the condition of \autoref{eq:cool_enhance} is satisfied. Above the green line, the condition of \autoref{eq:tcool_condition} is satisfied. As both conditions must be satisfied simultaneously to form a dense filament, the orange-shaded region represents the CGM parameter space to trigger the RSMBH-induced filament formation mechanism.
    \label{fig:cgm_condition}}
\end{figure}
%%%%%%%%%%

As demonstrated in \autoref{ssec:dependence_on_cgm_and_rsmbh}, our analytic model predicts the thermodynamic evolution of the CGM perturbed by the RSMBH precisely, and the formation of the RSMBH-induced filaments is independent of $M\sub{BH}$ and $v\sub{BH}$ when appropriate scales are considered. Using this model, we study the CGM parameters (density-temperature pairs) to form a dense filament with gravitational perturbation by an RSMBH in \autoref{fig:cgm_condition}. 

The parameter space relevant for the CGM surrounding the galaxy of interest is defined as follows: The galaxy has a stellar mass of $\sim 7 \times 10^9 \, \msun$ at $z = 1$ (\citetalias{vanDokkum2023a}). Galaxies with similar stellar mass are expected to develop into Milky Way-like galaxies at present \citep{Moster2018,Behroozi2019}. Motivated by results from hydrodynamical simulations studying such systems \citep[e.g.,][]{Oppenheimer2016,Tumlinson2017,Lochhaas2020}, we consider the two-dimensional parameter space of $n\sub{H} = [10^{-6} : 3 \times 10^{-4}] \, {\rm cm^{-3}}$ and $T = [10^4:10^6]$\,K in the analysis.

The blue line in \autoref{fig:cgm_condition} corresponds to $T\sub{peak}$ set by the condition of the enhancement of the cooling efficiency. If the temperature of the unperturbed CGM is above $T\sub{peak}$, \autoref{eq:cool_enhance} is never satisfied, as the perturbation increases the temperature of the CGM. So, the CGM temperature must be lower than the blue line prior to the RSMBH perturbation. The second condition that the cooling timescale must be shorter than the age of the observed Universe ($z = 1$) sets the lower limit of the CGM density. Above the green curve, the required condition (\autoref{eq:tcool_condition}) is satisfied.\footnote{When the net cooling time, $\tau\sub{cool}$, is compared to the time since the RSMBH ejection, the green line in \autoref{fig:cgm_condition} will be shifted to the right as higher CGM densities are required to form the filamentary structure in a shorter timescale.} As the gas evolves isobarically (\autoref{fig:profiles}), a critical pressure to ensure high enough density is set by \autoref{eq:tcool_condition}, and it is indicated by a diagonal line of $n\sub{H} \sim 2 \times 10^{-5} (T/2 \times 10^5 \, K)^{-1} \, {\rm cm^{-3}}$ in the figure. At $T > T\sub{peak}$, the cooling efficiency decreases with temperature and the critical density to satisfy \autoref{eq:tcool_condition} increases with temperature. Considering the two conditions simultaneously, we obtain the CGM parameter space to form dense filaments, indicated by the orange-shaded region.

%%%%%%%%%%%%%%%%%%%%%%%%%%%%%%%%%%%%%%%%%%%%%%%%%%
\section{Summary and Discussion}
\label{sec:summary}

Recent observations by \citet[][\citetalias{vanDokkum2023a}]{vanDokkum2023a} discovered a narrow linear feature in the CGM that extends $\sim 60$\,kpc from the centre of a star-forming galaxy at $z \sim 1$. The galaxy has an irregular morphology and is likely to have experienced recent mergers. \citetalias{vanDokkum2023a} interpreted the linear feature as a star-forming gas filament formed by a gravitational perturbation of an RSMBH ejected from the merging galaxies through interactions between SMBHs. 

This paper is the first theoretical attempt to test the RSMBH formation scenario for the thin, long filament in the CGM. We showed that, in principle, the strong gravity of the RSMBH can induce converging flows along the straight path of the RSMBH in the CGM. When the CGM temperature is below the peak temperature in the cooling function, $T\sub{peak}$, the gas compression increases the cooling efficiency. The cooling catastrophe along the RSMBH path then leads to the formation of a dense gas filament. Another requirement for the CGM to form dense gas filaments is that the timescale of radiative cooling is shorter than the age of the observed Universe and sets the lower limit for the CGM density. The RSMBH parameters (mass and velocity) determine the scale of the filament formation. Hydrodynamical simulations validate these expectations. We, therefore, conclude that the gravitational perturbation by RSMBHs is a viable channel for the formation of the observed linear feature and expect the CGM around the observed linear object to be warm ($T \la 2 \times 10^5$\,K) and dense ($n\sub{H} \ga 2 \times 10^{-5} (T/2 \times 10^5 \, K)^{-1} \, {\rm cm^{-3}}$).

\citetalias{vanDokkum2023a} suggested that the observed linear object might have formed in $\sim 40$\,Myr (since the RSMBH perturbs the CGM) if the RSMBH is located at the tip of the filament. On the other hand, the formation of a dense filament takes a few Gyr in our simulations assuming the CGM density of $n\sub{H}=10^{-4} \, {\rm cm^{-3}}$. Our analytical model explaining the simulation results predicts that having the CGM density of $n\sub{H} \sim 10^{-2} \, {\rm cm^{-3}}$ prior to the RSMBH perturbation, the filamentary structure could be formed in 40\,Myr. Such high-density environments would be in the tail of the CGM density distribution, explaining the rareness of the observed object. The gap might be filled by considering the higher ionisation fraction of the filament-forming gas. Suppose that feedback from nearby stars keeps the gas ionised. In that case, the cooling efficiency increases and consequently filament formation can be accelerated \citep{Gnedin2012}, while stellar feedback can delay or suppress filament formation. Self-consistent modelling of radiative cooling and heating processes would be an intriguing avenue for further exploration.

Another possible explanation for the timescale gap is that while the RSMBH leaves the `seed' to develop into the dense filament, it takes longer to have a high enough gas density to ignite star formation, as presented in our simulations. In that case, the RSMBH ran away from the vicinity of the host galaxy a few Gyrs ago. More detailed studies on the dynamical state of the gas and stars around the tip of the filament are needed to test this idea, if the RSMBH exists there. It would also help distinguish RSMBH-induced filament formation from the tidal arm/tail formation scenario \citep{Chen2023} as the latter expects that the SMBHs are located at the tip of the filament.

To estimate the baryonic mass involved in the converging flow along the path of the RSMBH, we use the formula:
%%%
\begin{equation}
M\sub{fil} = \pi \rho\sub{CGM} b\sub{comp}^2 L\sub{fil},
\end{equation}
%%%
where $\rho\sub{CGM}$ is the mass density of the background CGM, and $L\sub{fil} = v\sub{BH} \tau$ is the length of the filamentary structure at a time, $\tau$, since the RSMBH was ejected from its host galaxy. Using \autoref{eq:bcomp}, the value of $M\sub{fil}$ can be determined as:
%%%
\begin{equation}
    M\sub{fil} = 6.7 \times 10^4 \biggl ( \frac{n\sub{H}}{10^{-2}\,{\rm cm^{-3}}} \biggr ) \biggl ( \frac{M\sub{BH}}{10^7\,\msun} \biggr ) \biggl ( \frac{\tau}{40\,{\rm Myr}} \biggr )^2 \, \msun.
        \label{eq:mconv}
\end{equation}
%%%
Here, the mean molecular weight is assumed to be $0.6$. Assuming $M\sub{BH} \sim 2 \times 10^7 \, \msun$ and $\tau \sim 39 \, {\rm Myr}$, as suggested by \cite{vanDokkum2023a}, and $n\sub{H} \sim 10^{-2} \, {\rm cm^{-3}}$ as found in the argument of the timescale of filament formation, it can be estimated that $M\sub{fil} \sim 10^5 \, \msun$. If follow-up observations confirm the estimated mass of the filament, it will provide strong evidence in support of the RSMBH-induced filament formation scenario.

Some other potentially relevant processes are neglected in this paper for simplicity. For instance, pre-existing turbulent flows in the CGM can enhance or suppress the converging flows induced by the RSMBH, depending on the direction of the flows. As noted in \citetalias{vanDokkum2023a}, the observed feature is not perfectly linear and has a deflection, possibly reflecting the local kinematics of the CGM. Such fine structures may be reproduced by including pre-existing turbulent flows in simulations. Conversely, the observations of linear objects can constrain the local kinematics of the CGM if the RSMBH parameters are well-constrained. 

An important question for future studies is whether the RSMBH-induced mechanism eventually ignites star formation. Due to the resolution limit, the Jeans mass of the gas in the dense filament is $\sim 10^5 \, \msun$ at the end of run-A. While the Jeans mass is expected to decrease in the subsequent gas condensation, further exploration is needed. Although there is a rich literature investigating the fragmentation of the filament and the formation of stars \citep[e.g.,][]{Larson1985,Inutsuka1992,Burkert2004,Myers2017,Hoemann2023}, the outcome depends on the assumptions and detailed setup of the model. To reach a firm conclusion, numerical simulations with higher resolutions and setups motivated by observations of the linear object and its vicinity would be desired.

%%%%%%%%%%%%%%%%%%%%%%%%%%%%%%%%%%%%%%%%%%%%%%%%%%
\section*{Acknowledgements}
We thank our referee, Yannik Bah{\'e}, for careful reading and helpful comments, and Andreas Burkert, Hajime Fukushima, Oliver Hahn, Takanobu Kirihara, Yohei Miki, Masao Mori, Frank van den Bosch, Pieter van Dokkum, and Hidenobu Yajima for fruitful discussions. GO was supported by the National Key Research and Development Program of China (No. 2022YFA1602903) and the Fundamental Research Fund for Chinese Central Universities (Grant No. NZ2020021, No. 226-2022-00216). DN acknowledges support from the NSF AST-2206055 grant. We acknowledge the cosmology simulation database in the National Basic Science Data Center (NBSDC) and its funds, the NBSDC-DB-10.

%%%%%%%%%%%%%%%%%%%%%%%%%%%%%%%%%%%%%%%%%%%%%%%%%%
\section*{Data Availability}
The data and code underlying this article will be shared on reasonable request to the corresponding author.



\bibliographystyle{mnras}
\bibliography{rsmbh}

\begin{thebibliography}{}
\makeatletter
\relax
\def\mn@urlcharsother{\let\do\@makeother \do\$\do\&\do\#\do\^\do\_\do\%\do\~}
\def\mn@doi{\begingroup\mn@urlcharsother \@ifnextchar [ {\mn@doi@}
  {\mn@doi@[]}}
\def\mn@doi@[#1]#2{\def\@tempa{#1}\ifx\@tempa\@empty \href
  {http://dx.doi.org/#2} {doi:#2}\else \href {http://dx.doi.org/#2} {#1}\fi
  \endgroup}
\def\mn@eprint#1#2{\mn@eprint@#1:#2::\@nil}
\def\mn@eprint@arXiv#1{\href {http://arxiv.org/abs/#1} {{\tt arXiv:#1}}}
\def\mn@eprint@dblp#1{\href {http://dblp.uni-trier.de/rec/bibtex/#1.xml}
  {dblp:#1}}
\def\mn@eprint@#1:#2:#3:#4\@nil{\def\@tempa {#1}\def\@tempb {#2}\def\@tempc
  {#3}\ifx \@tempc \@empty \let \@tempc \@tempb \let \@tempb \@tempa \fi \ifx
  \@tempb \@empty \def\@tempb {arXiv}\fi \@ifundefined
  {mn@eprint@\@tempb}{\@tempb:\@tempc}{\expandafter \expandafter \csname
  mn@eprint@\@tempb\endcsname \expandafter{\@tempc}}}

\bibitem[\protect\citeauthoryear{{Baker}, {Centrella}, {Choi}, {Koppitz}, {van
  Meter}  \& {Miller}}{{Baker} et~al.}{2006}]{Baker2006}
{Baker} J.~G.,  {Centrella} J.,  {Choi} D.-I.,  {Koppitz} M.,  {van Meter}
  J.~R.,   {Miller} M.~C.,  2006, \mn@doi [\apjl] {10.1086/510448}, \href
  {https://ui.adsabs.harvard.edu/abs/2006ApJ...653L..93B} {653, L93}

\bibitem[\protect\citeauthoryear{{Behroozi}, {Wechsler}, {Hearin}  \&
  {Conroy}}{{Behroozi} et~al.}{2019}]{Behroozi2019}
{Behroozi} P.,  {Wechsler} R.~H.,  {Hearin} A.~P.,   {Conroy} C.,  2019,
  \mn@doi [\mnras] {10.1093/mnras/stz1182}, \href
  {https://ui.adsabs.harvard.edu/abs/2019MNRAS.488.3143B} {488, 3143}

\bibitem[\protect\citeauthoryear{{Bekenstein}}{{Bekenstein}}{1973}]{Bekenstein1973}
{Bekenstein} J.~D.,  1973, \mn@doi [\apj] {10.1086/152255}, \href
  {https://ui.adsabs.harvard.edu/abs/1973ApJ...183..657B} {183, 657}

\bibitem[\protect\citeauthoryear{{Binney} \& {Tremaine}}{{Binney} \&
  {Tremaine}}{2008}]{Binney2008}
{Binney} J.,  {Tremaine} S.,  2008, {Galactic Dynamics: Second Edition}.
Princeton University Press

\bibitem[\protect\citeauthoryear{{Bird}, {Ni}, {Di Matteo}, {Croft}, {Feng}  \&
  {Chen}}{{Bird} et~al.}{2022}]{Bird2022}
{Bird} S.,  {Ni} Y.,  {Di Matteo} T.,  {Croft} R.,  {Feng} Y.,   {Chen} N.,
  2022, \mn@doi [\mnras] {10.1093/mnras/stac648}, \href
  {https://ui.adsabs.harvard.edu/abs/2022MNRAS.512.3703B} {512, 3703}

\bibitem[\protect\citeauthoryear{{Burkert} \& {Hartmann}}{{Burkert} \&
  {Hartmann}}{2004}]{Burkert2004}
{Burkert} A.,  {Hartmann} L.,  2004, \mn@doi [\apj] {10.1086/424895}, \href
  {https://ui.adsabs.harvard.edu/abs/2004ApJ...616..288B} {616, 288}

\bibitem[\protect\citeauthoryear{{Campanelli}, {Lousto}, {Zlochower}  \&
  {Merritt}}{{Campanelli} et~al.}{2007}]{Campanelli2007}
{Campanelli} M.,  {Lousto} C.~O.,  {Zlochower} Y.,   {Merritt} D.,  2007,
  \mn@doi [\prl] {10.1103/PhysRevLett.98.231102}, \href
  {https://ui.adsabs.harvard.edu/abs/2007PhRvL..98w1102C} {98, 231102}

\bibitem[\protect\citeauthoryear{{Chen}, {LaChance}, {Ni}, {Di Matteo},
  {Croft}, {Natarajan}  \& {Bird}}{{Chen} et~al.}{2023}]{Chen2023}
{Chen} N.,  {LaChance} P.,  {Ni} Y.,  {Di Matteo} T.,  {Croft} R.,  {Natarajan}
  P.,   {Bird} S.,  2023, \mn@doi [\apjl] {10.3847/2041-8213/aced45}, \href
  {https://ui.adsabs.harvard.edu/abs/2023ApJ...954L...2C} {954, L2}

\bibitem[\protect\citeauthoryear{{Civano} et~al.,}{{Civano}
  et~al.}{2010}]{Civano2010}
{Civano} F.,  et~al., 2010, \mn@doi [\apj] {10.1088/0004-637X/717/1/209}, \href
  {https://ui.adsabs.harvard.edu/abs/2010ApJ...717..209C} {717, 209}

\bibitem[\protect\citeauthoryear{{Deane} et~al.,}{{Deane}
  et~al.}{2014}]{Deane2014}
{Deane} R.~P.,  et~al., 2014, \mn@doi [\nat] {10.1038/nature13454}, \href
  {https://ui.adsabs.harvard.edu/abs/2014Natur.511...57D} {511, 57}

\bibitem[\protect\citeauthoryear{{Escala}, {Larson}, {Coppi}  \&
  {Mardones}}{{Escala} et~al.}{2005}]{Escala2005}
{Escala} A.,  {Larson} R.~B.,  {Coppi} P.~S.,   {Mardones} D.,  2005, \mn@doi
  [\apj] {10.1086/431747}, \href
  {https://ui.adsabs.harvard.edu/abs/2005ApJ...630..152E} {630, 152}

\bibitem[\protect\citeauthoryear{{Gnedin} \& {Hollon}}{{Gnedin} \&
  {Hollon}}{2012}]{Gnedin2012}
{Gnedin} N.~Y.,  {Hollon} N.,  2012, \mn@doi [\apjs]
  {10.1088/0067-0049/202/2/13}, \href
  {https://ui.adsabs.harvard.edu/abs/2012ApJS..202...13G} {202, 13}

\bibitem[\protect\citeauthoryear{{G{\"u}ltekin} et~al.,}{{G{\"u}ltekin}
  et~al.}{2009}]{Gultekin2009}
{G{\"u}ltekin} K.,  et~al., 2009, \mn@doi [\apj] {10.1088/0004-637X/698/1/198},
  \href {https://ui.adsabs.harvard.edu/abs/2009ApJ...698..198G} {698, 198}

\bibitem[\protect\citeauthoryear{{Haehnelt}, {Davies}  \& {Rees}}{{Haehnelt}
  et~al.}{2006}]{Haehnelt2006}
{Haehnelt} M.~G.,  {Davies} M.~B.,   {Rees} M.~J.,  2006, \mn@doi [\mnras]
  {10.1111/j.1745-3933.2005.00124.x}, \href
  {https://ui.adsabs.harvard.edu/abs/2006MNRAS.366L..22H} {366, L22}

\bibitem[\protect\citeauthoryear{{Hoemann}, {Heigl}  \& {Burkert}}{{Hoemann}
  et~al.}{2023}]{Hoemann2023}
{Hoemann} E.,  {Heigl} S.,   {Burkert} A.,  2023, \mn@doi [\mnras]
  {10.1093/mnras/stad2517}, \href
  {https://ui.adsabs.harvard.edu/abs/2023MNRAS.tmp.2412H} {}

\bibitem[\protect\citeauthoryear{{Hoffman} \& {Loeb}}{{Hoffman} \&
  {Loeb}}{2007}]{Hoffman2007}
{Hoffman} L.,  {Loeb} A.,  2007, \mn@doi [\mnras]
  {10.1111/j.1365-2966.2007.11694.x}, \href
  {https://ui.adsabs.harvard.edu/abs/2007MNRAS.377..957H} {377, 957}

\bibitem[\protect\citeauthoryear{{Hopkins}}{{Hopkins}}{2013}]{Hopkins2013}
{Hopkins} P.~F.,  2013, \mn@doi [\mnras] {10.1093/mnras/sts210}, \href
  {https://ui.adsabs.harvard.edu/abs/2013MNRAS.428.2840H} {428, 2840}

\bibitem[\protect\citeauthoryear{{Hopkins}}{{Hopkins}}{2015}]{Hopkins2015}
{Hopkins} P.~F.,  2015, \mn@doi [\mnras] {10.1093/mnras/stv195}, \href
  {https://ui.adsabs.harvard.edu/abs/2015MNRAS.450...53H} {450, 53}

\bibitem[\protect\citeauthoryear{{Inutsuka} \& {Miyama}}{{Inutsuka} \&
  {Miyama}}{1992}]{Inutsuka1992}
{Inutsuka} S.-I.,  {Miyama} S.~M.,  1992, \mn@doi [\apj] {10.1086/171162},
  \href {https://ui.adsabs.harvard.edu/abs/1992ApJ...388..392I} {388, 392}

\bibitem[\protect\citeauthoryear{{Kalfountzou}, {Santos Lleo}  \&
  {Trichas}}{{Kalfountzou} et~al.}{2017}]{Kalfountzou2017}
{Kalfountzou} E.,  {Santos Lleo} M.,   {Trichas} M.,  2017, \mn@doi [\apjl]
  {10.3847/2041-8213/aa9b2d}, \href
  {https://ui.adsabs.harvard.edu/abs/2017ApJ...851L..15K} {851, L15}

\bibitem[\protect\citeauthoryear{{Kesden}, {Sperhake}  \& {Berti}}{{Kesden}
  et~al.}{2010}]{Kesden2010}
{Kesden} M.,  {Sperhake} U.,   {Berti} E.,  2010, \mn@doi [\apj]
  {10.1088/0004-637X/715/2/1006}, \href
  {https://ui.adsabs.harvard.edu/abs/2010ApJ...715.1006K} {715, 1006}

\bibitem[\protect\citeauthoryear{{Kitajima} \& {Inutsuka}}{{Kitajima} \&
  {Inutsuka}}{2023}]{Kitajima2023}
{Kitajima} K.,  {Inutsuka} S.-i.,  2023, \mn@doi [\apj]
  {10.3847/1538-4357/acb7ea}, \href
  {https://ui.adsabs.harvard.edu/abs/2023ApJ...945...39K} {945, 39}

\bibitem[\protect\citeauthoryear{{Kormendy} \& {Ho}}{{Kormendy} \&
  {Ho}}{2013}]{Kormendy2013}
{Kormendy} J.,  {Ho} L.~C.,  2013, \mn@doi [\araa]
  {10.1146/annurev-astro-082708-101811}, \href
  {https://ui.adsabs.harvard.edu/abs/2013ARA&A..51..511K} {51, 511}

\bibitem[\protect\citeauthoryear{{Larson}}{{Larson}}{1985}]{Larson1985}
{Larson} R.~B.,  1985, \mn@doi [\mnras] {10.1093/mnras/214.3.379}, \href
  {https://ui.adsabs.harvard.edu/abs/1985MNRAS.214..379L} {214, 379}

\bibitem[\protect\citeauthoryear{{Li} \& {Shi}}{{Li} \& {Shi}}{2021}]{Li2021}
{Li} G.-X.,  {Shi} X.,  2021, \mn@doi [\mnras] {10.1093/mnras/stab735}, \href
  {https://ui.adsabs.harvard.edu/abs/2021MNRAS.503.4466L} {503, 4466}

\bibitem[\protect\citeauthoryear{{Lochhaas}, {Bryan}, {Li}, {Li}  \&
  {Fielding}}{{Lochhaas} et~al.}{2020}]{Lochhaas2020}
{Lochhaas} C.,  {Bryan} G.~L.,  {Li} Y.,  {Li} M.,   {Fielding} D.,  2020,
  \mn@doi [\mnras] {10.1093/mnras/staa358}, \href
  {https://ui.adsabs.harvard.edu/abs/2020MNRAS.493.1461L} {493, 1461}

\bibitem[\protect\citeauthoryear{{Lotz}, {Jonsson}, {Cox}  \& {Primack}}{{Lotz}
  et~al.}{2008}]{Lotz2008}
{Lotz} J.~M.,  {Jonsson} P.,  {Cox} T.~J.,   {Primack} J.~R.,  2008, \mn@doi
  [\mnras] {10.1111/j.1365-2966.2008.14004.x}, \href
  {https://ui.adsabs.harvard.edu/abs/2008MNRAS.391.1137L} {391, 1137}

\bibitem[\protect\citeauthoryear{{Lotz}, {Dolag}, {Remus}  \& {Burkert}}{{Lotz}
  et~al.}{2021}]{Lotz2021}
{Lotz} M.,  {Dolag} K.,  {Remus} R.-S.,   {Burkert} A.,  2021, \mn@doi [\mnras]
  {10.1093/mnras/stab2037}, \href
  {https://ui.adsabs.harvard.edu/abs/2021MNRAS.506.4516L} {506, 4516}

\bibitem[\protect\citeauthoryear{{Lousto} \& {Zlochower}}{{Lousto} \&
  {Zlochower}}{2011}]{Lousto2011}
{Lousto} C.~O.,  {Zlochower} Y.,  2011, \mn@doi [\prl]
  {10.1103/PhysRevLett.107.231102}, \href
  {https://ui.adsabs.harvard.edu/abs/2011PhRvL.107w1102L} {107, 231102}

\bibitem[\protect\citeauthoryear{{Milosavljevi{\'c}} \&
  {Merritt}}{{Milosavljevi{\'c}} \& {Merritt}}{2001}]{Milosavljevic2001}
{Milosavljevi{\'c}} M.,  {Merritt} D.,  2001, \mn@doi [\apj] {10.1086/323830},
  \href {https://ui.adsabs.harvard.edu/abs/2001ApJ...563...34M} {563, 34}

\bibitem[\protect\citeauthoryear{{Moster}, {Naab}  \& {White}}{{Moster}
  et~al.}{2018}]{Moster2018}
{Moster} B.~P.,  {Naab} T.,   {White} S. D.~M.,  2018, \mn@doi [\mnras]
  {10.1093/mnras/sty655}, \href
  {https://ui.adsabs.harvard.edu/abs/2018MNRAS.477.1822M} {477, 1822}

\bibitem[\protect\citeauthoryear{{Myers}}{{Myers}}{2017}]{Myers2017}
{Myers} P.~C.,  2017, \mn@doi [\apj] {10.3847/1538-4357/aa5fa8}, \href
  {https://ui.adsabs.harvard.edu/abs/2017ApJ...838...10M} {838, 10}

\bibitem[\protect\citeauthoryear{{Ogiya}, {Hahn}, {Mingarelli}  \&
  {Volonteri}}{{Ogiya} et~al.}{2020}]{Ogiya2020}
{Ogiya} G.,  {Hahn} O.,  {Mingarelli} C. M.~F.,   {Volonteri} M.,  2020,
  \mn@doi [\mnras] {10.1093/mnras/staa444}, \href
  {https://ui.adsabs.harvard.edu/abs/2020MNRAS.493.3676O} {493, 3676}

\bibitem[\protect\citeauthoryear{{Oppenheimer} et~al.,}{{Oppenheimer}
  et~al.}{2016}]{Oppenheimer2016}
{Oppenheimer} B.~D.,  et~al., 2016, \mn@doi [\mnras] {10.1093/mnras/stw1066},
  \href {https://ui.adsabs.harvard.edu/abs/2016MNRAS.460.2157O} {460, 2157}

\bibitem[\protect\citeauthoryear{{Press}, {Teukolsky}, {Vetterling}  \&
  {Flannery}}{{Press} et~al.}{2002}]{Press2002}
{Press} W.~H.,  {Teukolsky} S.~A.,  {Vetterling} W.~T.,   {Flannery} B.~P.,
  2002, {Numerical recipes in C++ : the art of scientific computing}

\bibitem[\protect\citeauthoryear{{Saitoh} \& {Makino}}{{Saitoh} \&
  {Makino}}{2013}]{Saitoh2013}
{Saitoh} T.~R.,  {Makino} J.,  2013, \mn@doi [\apj]
  {10.1088/0004-637X/768/1/44}, \href
  {https://ui.adsabs.harvard.edu/abs/2013ApJ...768...44S} {768, 44}

\bibitem[\protect\citeauthoryear{{S{\'a}nchez Almeida}}{{S{\'a}nchez
  Almeida}}{2023}]{SanchezAlmeida2023b}
{S{\'a}nchez Almeida} J.,  2023, \mn@doi [\aap] {10.1051/0004-6361/202347098},
  \href {https://ui.adsabs.harvard.edu/abs/2023A&A...678A.118S} {678, A118}

\bibitem[\protect\citeauthoryear{{S{\'a}nchez Almeida}, {Montes}  \&
  {Trujillo}}{{S{\'a}nchez Almeida} et~al.}{2023}]{SanchezAlmeida2023a}
{S{\'a}nchez Almeida} J.,  {Montes} M.,   {Trujillo} I.,  2023, \mn@doi [\aap]
  {10.1051/0004-6361/202346430}, \href
  {https://ui.adsabs.harvard.edu/abs/2023A&A...673L...9S} {673, L9}

\bibitem[\protect\citeauthoryear{{Saslaw} \& {De Young}}{{Saslaw} \& {De
  Young}}{1972}]{Saslaw1972}
{Saslaw} W.~C.,  {De Young} D.~S.,  1972, \aplett, \href
  {https://ui.adsabs.harvard.edu/abs/1972ApL....11...87S} {11, 87}

\bibitem[\protect\citeauthoryear{{Saslaw}, {Valtonen}  \& {Aarseth}}{{Saslaw}
  et~al.}{1974}]{Saslaw1974}
{Saslaw} W.~C.,  {Valtonen} M.~J.,   {Aarseth} S.~J.,  1974, \mn@doi [\apj]
  {10.1086/152870}, \href
  {https://ui.adsabs.harvard.edu/abs/1974ApJ...190..253S} {190, 253}

\bibitem[\protect\citeauthoryear{{Sazonova} et~al.,}{{Sazonova}
  et~al.}{2021}]{Sazonova2021}
{Sazonova} E.,  et~al., 2021, \mn@doi [\apj] {10.3847/1538-4357/ac0f7f}, \href
  {https://ui.adsabs.harvard.edu/abs/2021ApJ...919..134S} {919, 134}

\bibitem[\protect\citeauthoryear{{Schure}, {Kosenko}, {Kaastra}, {Keppens}  \&
  {Vink}}{{Schure} et~al.}{2009}]{Schure2009}
{Schure} K.~M.,  {Kosenko} D.,  {Kaastra} J.~S.,  {Keppens} R.,   {Vink} J.,
  2009, \mn@doi [\aap] {10.1051/0004-6361/200912495}, \href
  {https://ui.adsabs.harvard.edu/abs/2009A&A...508..751S} {508, 751}

\bibitem[\protect\citeauthoryear{{Scott}, {Cortese}, {Lagos}, {Brinks},
  {Finoguenov}  \& {Coccato}}{{Scott} et~al.}{2022}]{Scott2022}
{Scott} T.~C.,  {Cortese} L.,  {Lagos} P.,  {Brinks} E.,  {Finoguenov} A.,
  {Coccato} L.,  2022, \mn@doi [\mnras] {10.1093/mnras/stac118}, \href
  {https://ui.adsabs.harvard.edu/abs/2022MNRAS.511..980S} {511, 980}

\bibitem[\protect\citeauthoryear{{Tanikawa} \& {Umemura}}{{Tanikawa} \&
  {Umemura}}{2011}]{Tanikawa2011}
{Tanikawa} A.,  {Umemura} M.,  2011, \mn@doi [\apjl]
  {10.1088/2041-8205/728/2/L31}, \href
  {https://ui.adsabs.harvard.edu/abs/2011ApJ...728L..31T} {728, L31}

\bibitem[\protect\citeauthoryear{{Townsend}}{{Townsend}}{2009}]{Townsend2009}
{Townsend} R.~H.~D.,  2009, \mn@doi [\apjs] {10.1088/0067-0049/181/2/391},
  \href {https://ui.adsabs.harvard.edu/abs/2009ApJS..181..391T} {181, 391}

\bibitem[\protect\citeauthoryear{{Tumlinson}, {Peeples}  \& {Werk}}{{Tumlinson}
  et~al.}{2017}]{Tumlinson2017}
{Tumlinson} J.,  {Peeples} M.~S.,   {Werk} J.~K.,  2017, \mn@doi [\araa]
  {10.1146/annurev-astro-091916-055240}, \href
  {https://ui.adsabs.harvard.edu/abs/2017ARA&A..55..389T} {55, 389}

\bibitem[\protect\citeauthoryear{{Van Wassenhove}, {Capelo}, {Volonteri},
  {Dotti}, {Bellovary}, {Mayer}  \& {Governato}}{{Van Wassenhove}
  et~al.}{2014}]{VanWassenhove2014}
{Van Wassenhove} S.,  {Capelo} P.~R.,  {Volonteri} M.,  {Dotti} M.,
  {Bellovary} J.~M.,  {Mayer} L.,   {Governato} F.,  2014, \mn@doi [\mnras]
  {10.1093/mnras/stu024}, \href
  {https://ui.adsabs.harvard.edu/abs/2014MNRAS.439..474V} {439, 474}

\bibitem[\protect\citeauthoryear{{Volonteri}, {Haardt}  \& {Madau}}{{Volonteri}
  et~al.}{2003}]{Volonteri2003}
{Volonteri} M.,  {Haardt} F.,   {Madau} P.,  2003, \mn@doi [\apj]
  {10.1086/344675}, \href
  {https://ui.adsabs.harvard.edu/abs/2003ApJ...582..559V} {582, 559}

\bibitem[\protect\citeauthoryear{{Wallin}, {Higdon}  \&
  {Staveley-Smith}}{{Wallin} et~al.}{1996}]{Wallin1996}
{Wallin} J.~F.,  {Higdon} J.~L.,   {Staveley-Smith} L.,  1996, \mn@doi [\apj]
  {10.1086/176920}, \href
  {https://ui.adsabs.harvard.edu/abs/1996ApJ...459..555W} {459, 555}

\bibitem[\protect\citeauthoryear{{Yagi}, {Komiyama}, {Yoshida}, {Furusawa},
  {Kashikawa}, {Koyama}  \& {Okamura}}{{Yagi} et~al.}{2007}]{Yagi2007}
{Yagi} M.,  {Komiyama} Y.,  {Yoshida} M.,  {Furusawa} H.,  {Kashikawa} N.,
  {Koyama} Y.,   {Okamura} S.,  2007, \mn@doi [\apj] {10.1086/512359}, \href
  {https://ui.adsabs.harvard.edu/abs/2007ApJ...660.1209Y} {660, 1209}

\bibitem[\protect\citeauthoryear{{Zaritsky} et~al.,}{{Zaritsky}
  et~al.}{2023}]{Zaritsky2023}
{Zaritsky} D.,  et~al., 2023, \mn@doi [\mnras] {10.1093/mnras/stad1964}, \href
  {https://ui.adsabs.harvard.edu/abs/2023MNRAS.524.1431Z} {524, 1431}

\bibitem[\protect\citeauthoryear{{de la Fuente Marcos} \& {de la Fuente
  Marcos}}{{de la Fuente Marcos} \& {de la Fuente
  Marcos}}{2008}]{de_la_Fuente_Marcos2008}
{de la Fuente Marcos} R.,  {de la Fuente Marcos} C.,  2008, \mn@doi [\apjl]
  {10.1086/587962}, \href
  {https://ui.adsabs.harvard.edu/abs/2008ApJ...677L..47D} {677, L47}

\bibitem[\protect\citeauthoryear{{van Dokkum}}{{van
  Dokkum}}{2023}]{vanDokkum2023b}
{van Dokkum} P.,  2023, \mn@doi [Research Notes of the American Astronomical
  Society] {10.3847/2515-5172/acd196}, \href
  {https://ui.adsabs.harvard.edu/abs/2023RNAAS...7...83V} {7, 83}

\bibitem[\protect\citeauthoryear{{van Dokkum} et~al.,}{{van Dokkum}
  et~al.}{2023}]{vanDokkum2023a}
{van Dokkum} P.,  et~al., 2023, \mn@doi [\apjl] {10.3847/2041-8213/acba86},
  \href {https://ui.adsabs.harvard.edu/abs/2023ApJ...946L..50V} {946, L50}

\makeatother
\end{thebibliography}

%%%%%%%%%%%%%%%%%%%%%%%%%%%%%%%%%%%%%%%%%%%%%%%%%%

%%%%%%%%%%%%%%%%% APPENDICES %%%%%%%%%%%%%%%%%%%%%

%\appendix

%%%%%%%%%%%%%%%%%%%%%%%%%%%%%%%%%%%%%%%%%%%%%%%%%%

% Don't change these lines
\bsp	% typesetting comment
\label{lastpage}
\end{document}